\documentstyle[11pt]{article}
\newcommand{\be}{\begin{equation}}
\newcommand{\en}{\end{equation}}
\newcommand{\bea}{\begin{eqnarray}}
\newcommand{\ena}{\end{eqnarray}}

\newcommand{\vs}[1]{\rule[ - #1 mm]{0 mm}{#1 mm}}
\newcommand{\hvs}[1]{\rule{0 mm}{#1 mm}}
\newcommand{\sm}[2]{\frac{\mbox{\footnotesize #1}\vs{-2}}
                   {\vs{-2}\mbox{\footnotesize #2}}}
\newcommand{\W}{{\sf W}}
\newcommand{\NP}[1]{Nucl.\ Phys.\ {\bf #1}}
\newcommand{\PL}[1]{Phys.\ Lett.\ {\bf #1}}

\newcommand{\CMP}[1]{Comm.\ Math.\ Phys.\ {\bf #1}}

\newcommand{\IJMP}[1]{Int.\ J.\ Mod.\ Phys.\ {\bf #1}}

\newcommand{\PRP}[1]{Phys.\ Rep.\ {\bf #1}}
\newcommand{\sect}[1]{\setcounter{equation}{0}\section{#1}}

\begin{document}
%
%
\renewcommand{\thefootnote}{\fnsymbol{footnote}}
\newpage
\setcounter{page}{0}
\pagestyle{empty}
%
%
\rightline{DFTT-40/93}
\rightline{hep-th/9307170}
\rightline{July 1993}

\vs{15}

\begin{center}
{
\LARGE {Classification of Structure Constants for \W-algebras from
Highest Weights}
 }\\[1cm]
{\large K.\ Hornfeck}\footnote{e-mail: HORNFECK@TO.INFN.IT; \hspace{0.5cm}
31890::HORNFECK}\\[0.5cm]
{\em INFN, Sezione di
Torino, Via Pietro Giuria 1, I-10125 Torino,
Italy}
\\[1cm]
\end{center}
\vs{15}

\centerline{ \bf{Abstract}}
We show that the structure constants of \W-algebras can be grouped
according to the lowest (bosonic) spin(s) of the algebra. The structure
constants in each group are described by a unique formula, depending on a
functional parameter $h(c)$ that is characteristic for each algebra. As
examples we give the structure constants ${\cal C}_{33}^4$ and ${\cal
C}_{44}^4$ for the algebras of type \W($2,3,4,\ldots$) (that include the
{\sf WA}$_{n-1}$-algebras) and the structure constant ${\cal C}_{44}^4$ for
the algebras of type \W($2,4,\ldots$), especially for all the algebras {\sf
WD}$_n$, {\sf WB}$(0,n)$, {\sf WB}$_n$ and {\sf WC}$_n$. It also
includes the bosonic projection of the super-Virasoro algebra and a yet
unexplained algebra of type \W($2,4,6$) found previously.
%
%
\renewcommand{\thefootnote}{\arabic{footnote}}
\setcounter{footnote}{0}
\newpage
\pagestyle{plain}
\sect{Introduction}
In recent years much effort has been put into the classification of
\W-algebras~\cite{FOR92,FRS92a,BW92}~\footnote{For a review on \W-algebras
see e.g.~ref.~\cite{BS93}.}. However, there seem still to be cases that
withstand such classifications. One example occurs for\linebreak
\W-algebras of type \W($2,4,6$) that show four solutions~\cite{KW91}: Two
have been identified with the algebras {\sf WB}$_3$ and {\sf WC}$_3$ and
a third solution is the bosonic projection of the super-Virasoro
algebra~\cite{Bou89,Kau91}. But a fourth algebra exists that has not yet
found an explanation~\cite{EHH93}. So far the only way to decide definitely
the number of solutions of a general \W-algebra of type \W($2,d_1, d_2,
\ldots$) is by explicit construction, a task that would go soon beyond the
available computer power.

In this paper we shall start from a different point of view that describes
the known \W-algebras based on the simple Lie-algebras but also
allows for additional solutions. Especially it includes the fourth
solution of \W($2, 4,6$) in a very surprising (though at the moment only
formal) way.

Computing structure constants by solving Jacobi-identites in the usual way
always makes use of the whole set of fields of the algebra. As a
consequence, even structure constants involving only low-dimensional
fields are impossible to obtain in the conventional way if the algebra
contains many or ``high''-dimensional fields. In this way already the
algebra of type \W($2,4,6$) is almost at the limit that can be accessed
with standard methods.

Contrary to that, our approach will only involve Jacobi-identities with
those fields that enter in the structure constants in which one is
interested. Starting from the highest weights~\cite{BPZ84} for the
\W-algebra we find that a set of highest weights exists that
considerably simplify the construction of the structure constants of the
algebra.

In section~2 we explain our philosophy in the case of the \W$_3$-algebra
and in the following sections~3 to~5 we apply our ideas to more
complicated \W-algebras and give some examples for structure constants
that will include the simplest ones for the algebras {\sf WA}$_{n-1}$, {\sf
WB}$_n$, {\sf WC}$_n$, {\sf WD}$_n$ and {\sf WB}($0,n$) as functions of
$n$.

\sect{Preliminaries: The \W$_3$-algebra}
The \W$_3$-algebra of Zamolodchikov~\cite{Zam85} is certainly one of the
best investigated \W-algebras. Because of its relative simplicity we shall
start our discussion with it.

A highest weight state $|h,w\!\!>$ is a primary state, i.e.
\bea
L_n \, |h,w\!\!> & = & 0 \hspace{2cm}\mbox{for $n > 0$}
\nonumber \\
L_o \, |h,w\!\!> & = & h \, |h,w\!\!> \label{primstate}
\ena
and is an eigenstate of $W_o$, whereas it is annihilated by the positive
frequency operators $W_n$:
\bea
W_n \, |h,w\!\!> & = & 0 \hspace{2cm} \mbox{for $n > 0$}  \nonumber \\
W_o \, |h,w\!\!> & = & w \,|h,w\!\!> \ \ . \label{hwcond}
\ena

In~\cite{FZ87} the degenerate representations have been investigated. By
applying the operators $L_n$ and~$W_n$ (with $n<0$) to the highest weight
states $|h,w\!\!>$ we can produce other primary states. At the first level
it is of the form
\be
\chi_1 \, = \, \left( \hvs{5}
2 h \, W_{-1} - 3 w \, L_{-1} \right) \, |h,w\!\!>
\label{p1}
\en
and at the second level
\be
\chi_2 \, = \, \left( \hvs{5} h \,(5h+1)\, W_{-2} - 12 w \, L^2_{-1} +
6 w \,(h+1)\, L_{-2}
\right) \, |h,w\!\!> \ \ .
\label{p2}
\en
For specific values of $h$ and $w$ these states become null-states and the
representation is degenerate. The state $\chi_1$ is a null-state whenever
\be
w^2 = - \frac{2 \,h^2 \,(c - 32 \, h - 2)}{9\, (5\,c+22)}
\label{n1cond}
\en
and $\chi_2$ is a null-state if $w$ is given by eq.~(\ref{n1cond}) and the
conformal dimension $h$ of the highest weight state obeys the quadratic
equation
\be
0 = \left(h + 1 - \sm{16}{3} \alpha_{+}^2\right)\,
\left(h+1-\sm{16}{3} \alpha_{-}^2\right) = h^2 +
\frac{h}{\vs{-2}\mbox{\footnotesize 18}} \, (c-14) +
\sm{1}{18}\, c         \ \ .
\label{n2cond}
\en
It follows that whenever $\chi_2$ is a null-state, also $\chi_1$ is
automatically a null-state (and the representation is completely
degenerate).

The state $|h,w\!\!>$ can be created by a field $P_o(z)$,
\be
|h,w\!\!> \, = \, \lim_{z\rightarrow 0} P_o(z) \, |0\!\!>
\en
and the highest-weight conditions~(\ref{primstate}) and~(\ref{hwcond})
translate into the condition that $P_o$ is a primary field of dimension
$h$ and in the operator product expansion (OPE) of $W$ with $P_o$ appears
as lowest-dimensional field $P_o$ itself with eigenvalue $w$, but also
higher-dimensional fields can occur,
\bea
T \star P_o & = & h \, P_o \nonumber \\
W \star P_o & = & w \, P_o + P_1 + P_2 \ \ ,
\label{opecompl}
\ena
where $P_1$ and $P_2$ are new primary fields of dimension $h+1$ and $h+2$,
respectively. They are no longer highest weights, since in the OPE of $W$
with $P_1$ and $P_2$ also $P_o$ will in general appear. These new fields
can be in some sense regarded to be connected to the states $\chi_1$ and
$\chi_2$. We expect therefore that in the case where $\chi_1$ and $\chi_2$
are null-fields, that is eqs.~(\ref{n1cond}) and~(\ref{n2cond}) are
satisfied, the OPEs~(\ref{opecompl}) reduce to~\footnote{However, using the
free-field realization for the \W$_3$-algebra and the vertex operator
$:\!\exp i \vec{\beta} \vec{\phi}\!:$ for $P_o$ (for details
see~\cite{FZ87}), one can see that this statement is not completely true.
Even when $\chi_1$ and $\chi_2$ are null-states, non-zero primary fields of
dimension $h+1$ and $h+2$ can be constructed. But the free-field
realization also justifies eq.~(\ref{hope}). On the other hand, in
principle the null-state conditions are not necessary conditions for the
OPEs~(\ref{hope}).}
\bea
T \star P_o & = & h \, P_o \nonumber \\
W \star P_o & = & w \, P_o                  \ \ .
\label{hope}
\ena

We can also make the ansatz~(\ref{hope}) to our starting point and ask
when it is consistent by evaluating the Jacobi-identity $(W, W, P_o)$. This
Jacobi-identity then reproduces exactly eqs.~(\ref{n1cond})
and~(\ref{n2cond})~\footnote{Remember that one Jacobi-identity produces in
general a set of constraints.}.

\sect{Adding more fields to \W$_3$}
We want to consider now a more general \W-algebra that has a spin-3 field
$W_3$, a spin-4 field $W_4$ and an arbitrary set of other primary fields
with dimension $>4$. We shall denote this algebra as \W($2,3,4,\ldots$) and
examples for such an algebra are the {\sf WA}$_{n-1}$-algebras. The general
OPEs of the low-dimensional fields are
\bea
W_3 \star W_3 & = & \sm{$c$}{3}\, I + \sqrt{\cal A}\, W_4 \nonumber \\
W_3 \star W_4 & = & \sm{3}{4}\, \sqrt{\cal A} \,W_3 + W_5 + W_{6,1}
\ \ .
\label{ope34}
\ena
The structure constant $\cal A$ depends on the actual algebra; we have not
assigned structure constants to $W_5$ and $W_{6,1}$, because in our
discussion will not enter their normalisation-even not their existence or
the number of spin-5 and spin-6 fields.

We are again interested in the case of the complete degenerate
representation, when the lowest-level states become null-states. Instead
of the ansatz~(\ref{hope}) we have now
\bea
W_3 \star P_o & = & w_3 \, P_o \nonumber \\
W_4 \star P_o & = & w_4 \, P_o + q_4 \, P_3 \label{34hope} \\
W_5 \star P_o & = & w_5 \, P_o + \ldots \ \ . \nonumber
\ena
Here $P_3$ is no longer a new field (as were $P_1$ and $P_2$ in the example
with the \W$_3$-algebra) but the composite primary field
\be
P_3 \, = \, \, :W_3 P_o: + \,w_3\,s_1(h,c)\,:T' P_o: + \,w_3 \,
s_2(h,c) \, :T P_o': + \,w_3 \, s_3(h,c) \,P_o'''
\label{compfield}
\en
that cannot be neglected from the OPE~(\ref{34hope}). Similary, the
``$\ldots$'' in eqs.~(\ref{34hope}) stand for other composite fields of the
same kind that do not enter the calculation up to the point we want to
carry out here, but would be needed for consistency. The Jacobi-identity
$(W_3, W_3, P_o)$ already fixes the eigenvalues $w_3$ and $w_4$ for any of
these algebras,
\bea
w_3^2 & = &
{{2\,{h^2}\,\left[ c\,(1 + 2\,h) + 2\,h\,(8\,{h} - 5)  \right] }\over
   {9\,\left[ c\,(3 - 2\, h) + 2\,h \right] }} \nonumber \\[2mm]
w_4 & = &
{{8\,\left( 2 + c \right) \,h\,
     \left[ c\,(1 + h) + 2 \, h\,( 9\,{h} - 7) \right] }\over
   {{\sqrt{\cal A}}\,\left( 22 + 5\,c \right) \,
     \left[ c\, (3- 2 \, h) + 2\,h \right] }} \ \ ,
\label{34eigen1}
\ena
and the Jacobi-identity $(W_3, W_4, P_o)$ yields $W_{6,1} = 0$ and
\bea
w_5 & = & {{48\,\left( 2 + c \right) \,\left[ c\,(1+h) + 9 \,h\,(2 \, h-7)
\right] \,
     \left[ c\,(3+2\,h) + 2\, h\,(32\,h - 27) \right] \,{ w_3}}\over
   {{\sqrt{A}}\,\left( 114 + 7\,c \right) \,
     \left[ c\,(3-2\,h) + 2\,h  \right] \,
     \left[ c\,(2\,h+1) + 2\, h\, (8 \,h - 5)  \right] }}
\label{34eigen2}
\\[2mm]
q_4 & = & {{144\,\left(h -1\right) \,{ w_3}}\over
   {{\sqrt{\cal A}}\,\left[ c\,(2\,h+1) + 2\,h\,(8\,h-5) \right] }}
\nonumber
\ena
and a third order equation for $h({\cal A},c)$.

On the other hand, $h({\cal A},c)$ can be inverted to give the structure
constant $\cal A$ as a function of the dimension $h$ of the highest weight
satisfying eq.~(\ref{34hope}),
\be
{\cal A}(h) \, = \,
{{256\,\left( 2 + c \right) \,\left[ c\,(h-2) - 2\,h  \right] \,
     \left[ c\,(h+1) +2\,h\,(9\,h - 7)  \right] }\over
   {\left( 22 + 5\,c \right) \,\left[ c\,(2\,h-3) - 2\,h  \right] \,
     \left[ c\,(2\,h+1) + 2\,h\,(8\,h-5) \right] }}
\label{Aofh}
\en
Including also the OPE
\be
W_4 \star W_4 \, = \, \sm{$c$}{4}\, I + {\cal B}\, \sqrt{\cal A}\, W_4 +
W_{6,2}
\en
we also find the form of the structure constant $\cal B$ from the
Jacobi-identity $(W_4, W_4, P_o)$:
\be
{\cal B}(h) \, = \,
{{ 3\,\left[ \hvs{4}
{c^3} \, (2\,{h^2} - 2\, h - 9) +
2 \,{c^2} \, (38\,{h^3} - 131\,{h^2}  + 75\,h - 3) -
4\,c \, h\, (45 \, h^2 + 17 \, h - 47 )
- 8 \, h^2 \, (41 h - 38)
\right] }\over
   {8\,\left( 2 + c \right) \,\left[ c\, (h -2) - 2\,h  \right] \,
     \left[ c\, (h+1) + 2\, h\, (9\, h - 7) \right] \vs{2}}}
\label{Bofh}
\en
The highest weight $h$ entering in these equations will be in general a
function of the central charge~$c$. Contrary to the \W$_3$-algebra (and
later to the algebras of type \W($2,5,\ldots$) and type \W($2,6,\ldots$))
it is not fixed by the Jacobi-identities and one could imagine that with an
arbitrary complicated function $h(c)$ one could produce arbitrary structure
constants $\cal A$ and $\cal B$. However, we believe that there are
additional strong constraints on the possible form of $h(c)$, without being
able to give an answer to that problem. If we would know the {\it general\/}
form of $h$, we could deduce the structure constant of any algebra of
type~\W($2,3,4,\ldots$), satisfying eq.~(\ref{34hope}).

For the {\sf WA}$_{n-1}$-algebras all the dimensions of the null-fields are
known~\cite{FL88} and especially the dimensions of the lowest level
null-fields satisfy the quadratic equation
\be
4 \, n^2 \, h^2 + 2\, h \, [ c -(n-1)\,(2\,n+1)] + c \, (n-1) \, = \, 0 \ \ .
\label{wasol12}
\en
When we denote by $h_{1,2}$ the two solutions of this equation, then ${\cal
A}(h_1) = {\cal A}(h_2)$ (as well as ${\cal B}(h_1) = {\cal B}(h_2)$) give
in a much easier way the structure constants for the {\sf
WA}$_{n-1}$-algebra, as obtained in~\cite{Hor92} using the free-field
realization of ref.~\cite{FL88}. In addition to $h_{1,2}$ there is a third
value of $h$ that reproduces the same structure constants and that is given
by
\be
h_3 \, = \, \frac{n+1}{2} \, \frac{c}{c+1-n}           \ \ .
\label{wasol3}
\en

But we know that the {\sf WA}$_{n-1}$-algebra is not the only possibility
for \W-algebras of this type. As has been shown in ref.~\cite{Hor93}, for
the algebra \W($2,3,4,5$) exist two different solutions: One is given by
{\sf WA}$_4$ and therefore described by eqs.~(\ref{wasol12})
and~(\ref{wasol3}). But also for the second algebra the structure constants
are of the form~(\ref{Aofh}) and~(\ref{Bofh}) and in this case the
dimensions of the highest weights satisfying eq.~(\ref{34hope}) are
\be
h_1 \, = \,- \sm{1}{2} \, \frac{c+4}{c+1} ;
\hspace{1.5cm} h_2 \, = \, \sm{1}{8} \, \frac{c \, (2-c)}{c+1};
\hspace{1.5cm} h_3 \, = \, \sm{3}{2}\, \frac{c}{c+1} \ \ .
\en

It would therefore be interesting to find other examples for algebras of
type \W($2,3,4,\ldots$) beyond {\sf WA}$_{n-1}$ to get a hint on the
general structure of $h$. But even with our limited knowledge we shall find
important consequences in the following examples.

\sect{\W-algebras of type \W($2,4,\ldots$)}
We shall now consider algebras, where the lowest-dimensional bosonic
primary field has spin~4, hence obeying the OPE
\be
W_4 \star W_4 \, = \, \sm{$c$}{4}\, I + \sqrt{\cal C}\, W_4 + W_6
\label{ope44}
\en
and that there is a highest weight $P_o$ with the property that
\bea
W_4 \star P_o & = & w_4 \, P_o \nonumber \\
W_6 \star P_o & = & w_6 \, P_o + \ldots        \ \ .
\label{4hope}
\ena
All information we are looking for is already contained in the
Jacobi-identity $(W_4, W_4, P_o)$; the eigenvalues $w_4$ and $w_6$ are
\bea
w_4 & = &- {{3\,h\,
{\left[ \hvs{4}
{c^2}\, (2 \, h^2 - 2\, h - 9) +
3\,c\, (12\, h^3 - 49\, h^2 + 40\, h - 2) +
2\,h\,(12 \,h^2 + 5 \, h - 14)
  \right] }
}\over
   {\left( 22 + 5\,c \right) \,{\sqrt{{\cal C}}}\,
     \left[ c \, (2\, h - 3)\, (h-2) + h\,(4\, h - 5)
  \right]
}}
\nonumber \\[2mm]
w_6 & = & {{2\,\left( c-1 \right) \,\left( 22 + 5\,c \right) \,h\,
     \left[ c \, (h + 2) + (5\, h - 2)\,(3\,h - 4) \right] \,
     \left[ c\, (2 \, h + 3) + 4\, h \, (12 \, h - 7) \right] }\over
   {3\,\left( 24 + c \right) \,\left(2\,c-1 \right) \,
     \left( 68 + 7\,c \right) \,
     \left[ c\,(2 \, h - 3)\,(h - 2) + h \, (4 \, h - 5)  \right]
}} \hspace{1cm}
\label{4eigen}
\ena
and the structure constant $\cal C$ is given by
\be
{\cal C}(h) \, = \,
{{36\,{{\left[ \hvs{4}
{c^2}\, (2 \, h^2 - 2\, h - 9) +
3\,c\, (12\, h^3 - 49\, h^2 + 40\, h - 2) +
2\,h\,(12 \,h^2 + 5 \, h - 14)
  \right] }^2}}\over
   {\left( 22 + 5\,c \right) \,\left[ c \, (h+1) + (3 \, h -1)\,(h-2)
\right]\,
     \left[ c \, (2\, h + 1) + 2\, h\, (8 \, h - 5) \right] \,
     \left[ c \, (2\, h - 3)\, (h-2) + h\,(4\, h - 5)
  \right]\vs{2} }}
\label{Cofh}
\en
One observation we make is the fact that ${\cal C} = 50/3$ for $c=1$,
independent of $h$ and hence independent of the actual algebra as long as
there is no pole at this value of the central charge.

We expect at least four different types of algebras of the form
\W($2,4,\ldots$), that are {\sf WD}$_n$, {\sf WB}($0,n$), {\sf WB}$_n$
and {\sf WC}$_n$~\footnote{Strictly speaking we should not include {\sf
WD}$_3$ and {\sf WD}$_4$ here since these algebras are not of the desired
type because they
have additional bosonic fields of dimension~3 and~4; it turns
out, however, that they are described as well by the formulas.}. For both
the algebras {\sf WD}$_n$ (integer $n\geq3$) and {\sf WB}($0,m$) with $n =
(2m+1)/2$ (half-integer $n\geq 5/2$) we have for the null-fields the
condition~\cite{FL89,FL90}
\bea
&&
4\, n\, (n-1)\, h_{1,2}^2 + h_{1,2}
\,(c + n\,(3-4n)) + (n-1)\,c \, = \, 0; \nonumber \\
&\mbox{equivalent to}\hspace{1cm} & h_3 \, = \, n   \ \ .
\label{wdsol}
\ena
In addition, for $n=3/2$ we find the bosonic projection of the
super-Virasoro algebra~\cite{Bou89,Kau91}.

We obtain therefore the structure constant ${\cal C}$ for {\sf WD}$_n$ and
{\sf WB}($0,n$) by simply setting $h=n$ or $h=(2n+1)/2$
into eq.~(\ref{Cofh}). We can compare the limit
$c\rightarrow \infty$ of the structure constant $\cal C$ of {\sf WD}$_n$
with its known classical limit, given in ref.~\cite{FRS92},
and find agreement.

For the non-simply laced algebras {\sf WB}$_n$ and {\sf WC}$_n$ the two
solutions $h_{\pm}$ of a quadratic equation no longer describe the same
algebra but interchange {\sf WB} and {\sf WC}~\cite{FKW92}.
The dimensions $h_1$ and
$h_2$ are now solutions of two different equations,
\bea
8 \, n\, (1 + n)\, h_1^2 + 2\,h_1\,(c - n\,(3 + 2\,n)) + c\,(2\,n-1) & = & 0
\label{wbcsol1} \\
(2\,n - 1)\,(2\,n + 1)\,h_2^2 + h_2 \,
(c - n\,(6\,n+1)) + c\,n + 2\,n^2 & = &
0
\label{wbcsol2}
\ena
and there is a third equation, corresponding to $h_3$
\be
2\,(c + 2\,n)\, h_3^2 - h_3
 \, (n\,(2\,n+ 3) + c\,(4\,n+3)) + c\,(n+1)\,(2\,n+1)
\, = \, 0                              \ \ ,
\label{wbcsol3}
\en
where $h_1$, $h_2$ and $h_3$ for the algebra {\sf WB} are given by the $(-,
+, +)$ solutions of eqs.~(\ref{wbcsol1}),
(\ref{wbcsol2})~and~(\ref{wbcsol3}), respectively, and those for the algebra
{\sf WC} by the solutions of the opposite sign.

The structure constants for {\sf WB}$_n$ and {\sf
WC}$_n$ are more complex and we shall therefore present them in the
appendix~A. Their classical limit for general $n$  coincide again with
the result of ref.~\cite{FRS92}. For $n=2$ and $n=3$ the result reproduces
the structure constants of {\sf WBC}$_2$~\cite{Bou88} and {\sf WB}$_3$ and
{\sf WC}$_3$~\cite{KW91}.

In all cases considered we can also ask for the limit $n \rightarrow
\infty$. We find the result that
\bea
{\cal C}[\mbox{\sf WD}_{\infty}] & = &
{\cal C}[\mbox{\sf WB}(0,\infty)] \, = \,
{\cal C}[\mbox{\sf WB}_{\infty}] \, = \, 54\, \frac{(3\,c+2)^2}{(c+2)\,(5\,
c+22)}
 \label{wblim} \\
{\cal C}[\mbox{\sf WC}_{\infty}] & = & 2 \,\frac{(19\,c-34)^2}{(2\,c-1)\,(5\,
c+22)}                          \ \ .
\label{wclim}
\ena
The even-dimensional fields of the algebra \W$_\infty$~\cite{PRS90}
form a subalgebra
and it turns out that for this subalgebra the structure constant $({\cal
C}_{44}^4)^2$ is identical to eq.~(\ref{wblim}) after making a suitable
change of basis that brings us from the primary fields used here to the
fields where the \mbox{\W$_\infty$-algebra} linearizes. Since also the
values for $w_4$ and $w_6$ for \W$_\infty$ and {\sf WB}$_\infty$ agree, we
conjecture that {\sf WB}$_\infty$ is the subalgebra of even-dimensional
fields of \W$_\infty$ (that is equivalent to {\sf WA}$_\infty$). In the
same way the algebra {\sf WC}$_\infty$ is the subalgebra of the
even-dimensional fields of \W$_{1 + \infty}$, constructed in
ref.~\cite{PRS90b}. It is also interesting to note that these limits can
also be obtained by inserting $h=1$ or $h = 1/2$, respectively, into
eq.~(\ref{Cofh}) as these values are the solutions to eq.~(\ref{wbcsol2})
in the $n\rightarrow \infty$ limit.

So far we have taken in all solutions $n$ to be positive. However, from our
point of view there is no restriction on the sign of $n$ and we could as
well imagine negative values: If the Jacobi-identities are satisfied for
any positive $n$ they will also be satisfied for negative~$n$, as long as
there is no structure constant that diverges in this case. The only open
question will then be, whether the algebra closes with a finite number of
fields.

Indeed, there exists at least one example of
an algebra that can be described by a negative~$n$: This is the
``strange'' solution of the algebra \W($2,4,6$)~\cite{KW91,EHH93}. One
obtains the structure constants of this
algebra by setting $h = n=-1$. Very formally we can therefore
say that this algebra is~{\sf WD}$_{-1}$. Whether one can give to this
formal notation a mathematical meaning has to be seen. The
structure constant $\cal C$ does not diverge in this case in the limit $c
\rightarrow \infty$ because $n$ is negative, but because
$n$ takes one of four special values where $\cal C$ diverges
in this limit (another is e.g.\ $n = 3/2$, the bosonic projection
of the super-Virasoro-algebra).

Minimal models for the algebra {\sf WD}$_n$ are given by
\be
c(n) \, = \, n - 2 \, n \, (n-1) \,(2n-1 )\, \left(\frac{(p-q)^2}{p\,q}\right)
\ \ .
\label{ratmod}
\en
When we take our interpretation for the fourth solution of \W($2,4,6$) as
{\sf WD}$_{-1}$ more seriously, then we should expect minimal models for this
algebra, following eq.~(\ref{ratmod}) for the central charge, to be a
sub-set of
\be
c(-1) = -1 + 12 \left( \frac{(p-q)^2}{p\,q} \right) \ \ .
\en
Indeed, all examples for minimal models found in~\cite{EHH93} obey this
form with $q = p+1$.

Even more intrigueing, we can also give a formula for the dimension $h$ in
this case that is obtained by taking $n=-1$ for the highest weights of~{\sf
WD}$_n$: Let
\be
x(\vec{l}_i) \, = \, l_i \, (p+1) - l'_i \, p \hspace{1.5cm} \mbox{with}
\hspace{0.5cm} \vec{l}_i  \, = \, (l_i,l'_i), \hspace{1.5cm}i=1,2
\en
then the dimensions of the highest weights can be written as
\be
h(\vec{l}_1, \vec{l}_2; r,r') \, = \,
\frac{ r \,\left( x(\vec{l}_1) - 1\right) \left( x(\vec{l}_1) +
2 \,x(\vec{l}_2) - 6 - r
\right) + r'\, \left( x(\vec{l}_2) - 1\right) \left( x(\vec{l}_2) - 4 -
r'\right)}{2\,p\,(p+1)}
\label{hwofd}
\en
where $p = 2  m + 1$ with positive integer $m$ and
\bea
l_2 & = & l_1 \, = \, (1,1) \hspace{2cm} r=r'=1\\[2mm]
\mbox{or}\hspace{0.8cm} l_2 & = & (1,1) \hspace{1cm}
l_1 \,= \, (1,2), (2,1), (2,2) \hspace{1.2cm}1 \leq r \leq m-1,
\hspace{0.5cm}r'=1  \\[2mm]
\mbox{or}\hspace{0.8cm}
l_2 & = & (1,2) \hspace{1cm} l_1 \, = \, (2,1) \nonumber \\
l_2 & = & (2,1) \hspace{1cm} l_1 \, = \, (1,2) \hspace{2cm}
1 \leq r < r' \leq m-1
\ena
The $m^2$ dimensions~(\ref{hwofd})
reproduce the values for the examples given in ref.~\cite{EFH92}
for~$m=3$ and in ref.~\cite{EHH93}
for~$m=4$ and~$m=5$. For $m=2$ we get the highest weights of the Virasoro
minimal model at $c=-3/5$. However, the restrictions used here ($q=p+1$,
$p$ odd, bounds for $\vec{l}_i$ and $r,r'$) were found completely
by comparing with these examples~\footnote{In a ``normal'' {\sf WD} minimal
model the r\^{o}les of $l,l'$ and $r,r'$ are interchanged: There, $r$ and
$r'$ run over a fixed range, whereas the bounds for $l$ and $l'$ depend on
$p$.}; until we have a better understanding what
we mean by ``{\sf WD}$_{-1}$'', these bounds remain mysterious.

We presented here this formal identification only for the structure
constant ${\cal C}_{44}^4$, but indeed we inluded also the spin-6 field and
computed all structure constants up to this field (see appendix~B) and they
agree with this identification (of course this is also true for the other
\W($2,4,6$)-algebras).

It might also be interesting to note that with the ansatz that $h_{1,2}$
are given by both the solutions $h_{\pm}$ of a general quadratic equation
of the form
\be
h^2 + h \, (a_1 c + a_2) + (a_3 c + a_4) \, = \, 0
\label{ansatzqu}
\en
with free parameters $a_1, \ldots, a_4$, the algebra \W($2,3,4,\ldots$)
has only one solution of this type, given by {\sf WA}$_{n-1}$,
eq.~(\ref{wasol12}), whereas for \W($2,4,\ldots$) we find apart from the
solution of {\sf WD}$_n$ and {\sf WB}($0,m$), eq.~(\ref{wdsol}), and the
special case {\sf WB}$_2$={\sf WC}$_2$ another possibility, given by
\bea
&&4\,n\,(n-1)\, h_{1,2}^2 + h_{1,2}\, (c + (n-1)\,(1 - 4n)) +(n-1)\, c +
(n-1)^2 \, = \, 0; \nonumber \\[1mm]
&& h_3 \, = \, n \, \frac{c}{c+n-1}                      \ \ .
\label{wxsol}
\ena
We do not know of any example of an \W-algebra connected to this case and
is has to be seen, whether such an algebra exists (for some or any value of
$n$).

\sect{\W-algebras of type \W($2,5,\ldots$) and type \W($2,6,\ldots$)}
For the algebras of type \W($2,5,\ldots$) and \W($2,6,\ldots$) the
Jacobi-identities become strong enough to determine also the dimension of
the heighest weights. Among the solutions for \W($2,5,\ldots$) we would
expect the algebra~{\sf WE}$_6$ = \W($2,5,6,8,9,12$). For the type \W($2,6,
\ldots$) we should find the orbifold of \W$_3$ (because of the
parity-conservation of \W$_3$, the spin-6 composite field $W_6 \sim \,:
W_3 W_3:\, + \ldots$ spans an algebra of the desired type and for the same
reason $W_3$ plays the r\^{o}le of $P_o$), {\sf WG}$_2$ = \W($2,6$), {\sf
WF}$_4$ = \W($2,6,8,12$) and {\sf WE}$_7$ = \W($2,6,8,10,12,14,18$).

In the case of \W($2,5,\ldots$) in the simplest Jacobi-identity $(W_5, W_5,
P_o)$ does not enter a structure constant and we find a $5^{\mbox{\tiny
th}}$-order condition for $h(c)$. For \W($2,6,\ldots$) we find the
structure constant ${\cal C}_{66}^6(h)$ and the solutions for~$h$
\bea
h = 3 &\mbox{(for the orbifold of \W$_3$);} \\
28 \, h^2 + h \, (c - 26) + 3 \, (c+2) \, = \, 0 & \mbox{leading to
the structure
constant of {\sf WG}$_2$,} \nonumber \\
& \hspace{-3cm}\mbox{ as given in~\cite{FS90,BFK91,KW91};}
\ena
as well as a third order equation,
\be
88 + 20\,c - 64\,h - 13\,c\,h - 22\,{h^2} + 2\,c\,{h^2} + 16\,{h^3}\, = \, 0
\en
and a $6^{\mbox{\tiny th}}$-order condition.

However, it is doubtful that the {\sf WE}-algebras are contained in these
solutions for \W($2,5,\dots$) and \W($2,6,\ldots$), since the algebras {\sf
WE}$_i$ are simply laced we would therefore expect that $h$ again satisfies
a quadratic equation. Also the algebra {\sf WF}$_4$ is missing: There is no
spin-10 field in {\sf WF}$_4$, but only our
solution corresponding to {\sf WG}$_2$ is without it.

We have also to remember that we are only looking at one Jacobi-identity
(($W_5, W_5, P_o)$ or $(W_6, W_6, P_o)$, respectively) at this point and
we therefore get necessary conditions for $h$. It is not at all evident
that the solutions to the $5^{\mbox{\tiny th}}$-order condition in the case
of \W($2,5,\ldots$) and the $3^{\mbox{\tiny rd}}$-order and the
$6^{\mbox{\tiny th}}$-order condition for \W($2,6,\ldots$) lead to
\W-algebras.

One reason why {\sf WF}$_4$ (and {\sf WE}$_i$?) are missing might be that
the condition that {\em all\/} the fields $P_i$, $i \neq 0 $, drop out of the
OPE of the $W$-fields with $P_o$ {\em simultaneously\/} is no longer
satisfied.

\newpage
\sect{Conclusions}
All calculations are based on the assumption that there exists a highest
weight $P_o$ such that its OPE with the lowest-dimensional bosonic
generator $W$ obeys the simple rule $W \star P_o =  w \,P_o$.
When we consider as an example the {\sf WA}$_{n-1}$-algebra, the fields with
even parity form a subalgebra of the type \W($2,4,\ldots$). However, as it
is clear from eq.~(\ref{34hope}), the field $P_3$ given by
eq.~(\ref{compfield}) is not of the form we need for null-states and
therefore a reduction to eq.~(\ref{4hope}) will only happen
accidentally. By taking the solution e.g.\ for {\sf WA}$_{n-1}$, ($n \geq
3$ integer), this condition is only satisfied for $n = 4$ and $n =
\infty$. We see that our simple ansatz will not be be true in general for
orbifold constructions except single cases (see examples given).

Moreover, instead of our original ansatz we could think of a different
one of the form $W \star P_o = w \, P_1$ (or other fields than $P_o$).
Such an ansatz would be automatically needed for example if $W$ is
fermionic. We have not studied whether in the cases considered here a
different ansatz would lead to different solutions. In some examples we
checked that it reproduced known results (with different values of $h$).
However, it seems likely that for the algebras {\sf WE} and {\sf WF} we
need a even more complicated ansatz.

Another draw-back of our approach is that we are not able to deduce the
field content of our algebras. This has to be obtained by other means. On
the other hand, we were able to obtain structure constants for a variety of
algebras, like {\sf WD}$_n$ or {\sf WB}$_n$ as functions of $n$, a task
that would be principally impossible by evaluating the Jacobi-identities of
the algebra itself.

Moreover it is no longer necessary to give to each single algebra a new set
of solution of the structure constants, but it is sufficient to write down
the general function ${\cal C}(h)$ and then for each algebra once its form
of $h(c)$ (this is especially true for \W-algebras of the same type).

The calculations up to the point presented are relatively simple. We could
also get higher structure constants in the same way, but then calculations
will turn out to be more cumbersome because then also composite primary
fields of the algebra will contribute (that did not enter so far).

Of course also other examples can be considered, for example algebras with
a set of several \mbox{spin-4} fields etc. It would also be interesting to
find other examples that correspond to negative~$n$ in eq.~(\ref{wdsol}).
\vs{20}

{\large Acknowledgements}

Calculations have been carried out with the Mathematica-package for
calculating operator product expansions by K.~Thielemans~\cite{Thi91}.

I would like to thank W.~Eholzer, A.~Honecker, R.~H\"{u}bel and E.~Ragoucy
for helpful discussions and comments on the manuscript.

I am grateful to the University of Torino for kind hospitality during this
work and to the Laboratoire de Physique Th\'{e}orique ENSLAPP, Groupe
d'Annecy, during its final stage.

\newpage

\appendix

\sect{Structure constant for {\sf WB}$_n$ and {\sf WC}$_n$}
The structure constant $\cal C$ (see eq.~(\ref{ope44})) for {\sf WB}$_n$
and {\sf WC}$_n$ has the general form
\be
{\cal C}[\mbox{\sf WB}_n, \mbox{\sf WC}_n] \,\, = \,\, \frac{g_1 \pm g_2
\sqrt{y}}{g_3}
\label{CofBC}
\en
where
\bea
g_1 & = &
{c^6}\,\left( -1694 + 1191\,n + 2202\,{n^2} - 712\,{n^3} - 672\,{n^4} +
     96\,{n^5} + 64\,{n^6} \right) \, + \nonumber \\
&&  3\,{c^5}\,\left( 110 - 12167\,n - 9349\,{n^2} + 20942\,{n^3} +
     4816\,{n^4} - 8920\,{n^5} - 432\,{n^6} + 960\,{n^7} \right)\,  +
    \nonumber \\
&& 3\,{c^4}\,
   \left( 1024 + 10626\,n - 68841\,{n^2} - 193562\,{n^3} + 32218\,{n^4} +
     163136\,{n^5} - 58656\,{n^6} \right. \nonumber \\
&&\hspace{1.5cm} \left.- 32672\,{n^7} + 12704\,{n^8} \right)  \, +
     \nonumber \\
&&  2\,{c^3}\,\left( -2204 + 8274\,n + 225492\,{n^2} + 384408\,{n^3} -
     866403\,{n^4} - 1414656\,{n^5} + 469512\,{n^6} \right. \nonumber \\
&&\hspace{1.5cm} \left. + 436560\,{n^7} -
     340080\,{n^8} + 54208\,{n^9} \right) \, +  \\
&&  12\,{c^2}\,n\,\left( -3118 - 21773\,n + 8646\,{n^2} + 314624\,{n^3} +
     318922\,{n^4} - 377096\,{n^5}\right. \nonumber \\
&&\hspace{1.5cm} \left.  - 345264\,{n^6} + 77088\,{n^7} -
     1760\,{n^8} \right) \, + \nonumber \\
&&  24\,c\,{n^2}\,\left( -1260 - 13824\,n - 73493\,{n^2} + 3500\,{n^3} +
     269320\,{n^4} + 199120\,{n^5} - 14800\,{n^6} - 8192\,{n^7} \right) \, +
     \nonumber \\
&&  32\,{n^3}\,\left( -1960 - 6090\,n - 39153\,{n^2} - 88436\,{n^3} -
     56328\,{n^4} + 2256\,{n^5} + 8816\,{n^6} \right)  \nonumber \\[2mm]
g_2 & = & \left( n-2  \right) \,
\left( 1 - c \right) \,{{\left( 22 + 5\,c \right) }^2}\,
  \left( 5\,c + 28\,n + 2\,c\,n + 40\,{n^2} \right) \,
  \left( c + 4\,c\,n + 6\,{n^2} + 8\,{n^3} + 8\,{n^4} \right) \\[2mm]
g_3 & = & \left(   n-1 \right) \,
\left( -1 + 2\,c \right) \,\left( 22 + 5\,c \right) \,
  \left( 2\,c - 5\,n + c\,n + 2\,{n^2} \right) \,
  \left( 2 + c + 11\,n + c\,n + 12\,{n^2} \right) \, \nonumber \\
&&  \left( 3\,c + 10\,n + 2\,c\,n + 12\,{n^2} \right) \,
  \left( 3\,c + 20\,n - 8\,c\,n + 8\,{n^2} + 4\,c\,{n^2} \right) \\[2mm]
\mbox{and}&& \nonumber \\
y & = &
c^2 + 2\,c\,n\,(1 - 6\,n - 8\,n^2) + n^2\,(3 + 2\,n)^2    \ \ .
\ena

For the algebra {\sf WB}$_n$ we have to take the ``$-$'', for the algebra
{\sf WC}$_n$ the ``$+$''-sign in~(\ref{CofBC}).

\sect{Structure constants for \W($2,4,6,\ldots$)}
For the algebra with OPEs
\bea
W_4 \star W_4 & = & \sm{$c$}{4}\, I + \sqrt{\cal C}\, W_4 + \sqrt{\cal D}
\,W_6 \nonumber \\
W_4 \star W_6 & = &  \sm{2}{3} \sqrt{\cal D} \, W_4 + {\cal E} \sqrt{\cal
C} \, W_6 + W_{8,1} \\
W_6 \star W_6 & = & \sm{$c$}{6} \, I + \sm{2}{3} {\cal E}\sqrt{\cal C} \,
W_4 + {\cal F}\sqrt{\cal D} \, W_6 + W_{8,2} + W_{10} \nonumber
\ena
we find in addition to
the structure constant $\cal C$, eq.~(\ref{Cofh}), the
structure constants
\bea
{\cal D}(h) & = & 12 \,{{\left(c - 1 \right) \,{{\left( 22 + 5\,c \right) }^2}
\,
     \left[ c \, (h + 2) + (5\,h-2)\,(3\,h - 4) \right] \,}  \over
   {\left( 24 + c \right) \,\left( 2\,c - 1\right) \,
     \left( 68 + 7\,c \right) \,
     \left[ c \, (h + 1) + (3\, h -1)\, (h-2) \right] }}\, \times \nonumber \\
&&\hspace{1cm} {{     \left[ c \, (2\, h + 3) + 4\, h \, (12\, h -7)
  \right] \,
     \left[ c\, (2\, h - 5)\,(h - 3) + 2\, h\,(4\, h - 7)
  \right] }\over
{     \left[ c\, (2\, h + 1) + 2\, h\, (8 h - 5) \right] \,
     \left[ c\, (2\, h - 3)\, (h - 2) + h \, (4 \, h - 5)  \right] }} \\[1mm]
{\cal E}(h) & = & \nonumber \\
&& \hspace{-1 cm}
{{\left(  2\,c - 1 \right) \,\left( 68 + 7\,c \right) \,
\left( h - 4  \right) \,\left[ c\, (h + 3) + 4\, h \,(12 \, h - 13)
 \right]} \over
{6\,\left( 24 + c \right)\,
{\left[ \hvs{4}
{c^2}\, (2 \, h^2 - 2\, h - 9) +
3\,c\, (12\, h^3 - 49\, h^2 + 40\, h - 2) +
2\,h\,(12 \,h^2 + 5 \, h - 14)
  \right] }
} } \\[2mm]
{\cal F}(h) & = & \sm{5}{9}
{{1}\over
   {\left(  c- 1  \right) \,\left( 24 + c \right) \,
     \left( 22 + 5\,c \right)}} \, \times \nonumber \\
&& {{1} \over{    \left[ c\, (h + 2) + (5\, h - 2)\,(3\, h -4)
 \right] \,
     \left[ c\, (2\, h + 3) + 4\, h \, (12 \, h - 7) \right] \,
     \left[ c\, (2\, h - 5)\,(h - 3) + 2\, h \, (4\, h - 7) \right] }}
\, \times
\nonumber \\
&&\left\{ \hvs{5} 4{c^6}\,\left( -9 - h + {h^2} \right) \,
   \left( -87 - 4\,h + 4\,{h^2} \right)    + \right.\nonumber \\
&&  3\,{c^5}\,\left( 17016 - 32683\,h + 36363\,{h^2} - 2692\,{h^3} -
     3732\,{h^4} + 656\,{h^5} \right)       + \nonumber \\
&&  3\,{c^4}\,\left( 10488 - 537784\,h + 912383\,{h^2} - 779183\,{h^3} +
     510680\,{h^4} - 155788\,{h^5} + 16704\,{h^6} \right)  + \nonumber \\
&&  24\,{c^3}\,\left( -3900 + 39820\,h + 567630\,{h^2} - 1411325\,{h^3} +
     1089288\,{h^4} - 319940\,{h^5} + 34032\,{h^6} \right) + \nonumber \\
&&  24\,{c^2}\,\left( -41856 + 100768\,h - 554146\,{h^2} + 1010616\,{h^3} -
     696919\,{h^4} + 127816\,{h^5} + 20976\,{h^6} \right)  + \nonumber \\
&&  96\,c\,h\,
   \left( 113632 - 496572\,h + 799600\,{h^2} - 505275\,{h^3} +
     124168\,{h^4} - 15600\,{h^5} \right)
+                           \nonumber \\
&& \left. 1024\,{h^2}\,\left( -10759 + 53477\,h - 94159\,{h^2}
+ 66156\,{h^3} -
    15696\,{h^4} \right)  \hvs{5} \right\}                    \ \ .
\ena

For the algebra \W($2,4,6$) this is the complete set of structure constants
among the simple fields. Especially they agree with the ones given
in~\cite{KW91} for the bosonic projection of the super-Virasoro algebra for
$h=n=3/2$ and with the special fourth solution for $h=n=-1$.

\newpage
{

}
\end{document}